\documentclass[conference]{IEEEtran}
\usepackage{graphicx}
\usepackage{subfigure}
\usepackage{amsmath}
\usepackage{color}
\usepackage{cite}
\usepackage{amsmath,amssymb,amsfonts}
\usepackage{algorithmic}
\usepackage{textcomp}
\usepackage{booktabs}
\usepackage{epstopdf}
\usepackage{array}
\usepackage{pgffor}

\renewcommand{\v}[1]{\mbox{$\mathbf{#1}$}}

\newcommand{\vg}[1]{\mbox{$\boldsymbol{#1}$}}

\newcommand{\fig}[1]{\mbox{Figure~\ref{#1}}}
\newcommand{\eq}[1]{\mbox{Eq.~(\ref{#1})}}

\begin{document}

\title{Study on a Poisson's Equation Solver Based On Deep Learning Technique}

\author{\IEEEauthorblockN{Tao Shan,Wei Tang, Xunwang Dang, Maokun Li, Fan Yang, Shenheng Xu, and Ji Wu}
\IEEEauthorblockA{Tsinghua National Laboratory for Information Science and Technology (TNList),\\
Department Of Electronic Engineering, Tsinghua University, Beijing, China\\
Email: maokunli@tsinghua.edu.cn}
}

\maketitle

\begin{abstract}
In this work, we investigated the feasibility of applying deep learning techniques to solve  Poisson's equation. A deep convolutional neural network is set up to predict the distribution of electric potential in 2D or 3D cases. With proper training data generated from a finite difference solver, the strong approximation capability of the deep convolutional neural network allows it to make correct prediction given information of the source and distribution of permittivity. With applications of L2 regularization, numerical experiments show that the predication error of 2D cases can reach below 1.5\% and the predication of 3D cases can reach below 3\%, with a significant reduction in CPU time compared with the traditional solver based on finite difference methods.
\end{abstract}
\IEEEpeerreviewmaketitle

\section{Introduction}
Computational electromagnetic simulation has been widely used in research and engineering, such as antenna and circuit design, target detection, geophysical exploration, nano-optics, and many other related areas \cite{Chew2001}. Computational electromagnetic algorithms serve as the kernel of simulation. They solve Maxwell's equations under various materials and different boundary conditions. In these algorithms, the domain of simulation is usually discretized into small subdomains and the partial differential equations is converted from continuous into discrete forms, usually as matrix equations. These matrix equations are either solved using direct solvers like LU decomposition, or iterative solvers such as the conjugate gradient method \cite{Golub1996}. Typical methods in computational electromagnetics include finite difference method (FDM) \cite{Taflove2005}, finite element method (FEM) \cite{Jin2014}, method of moments (MOM) \cite{Harrington1993}, and etc. Practical models are usually partitioned into thousands or millions of subdomains, and matrix equations with millions of unknowns are solved on computers. This usually requires a large amount of CPU time and memory. Therefore, it is still very challenging to use full-wave computational electromagnetic solvers to applications that require real-time responses, such as radar imaging, biomedical monitoring, fluid detection, non-destructive testing, etc. The speed of electromagnetic simulation still cannot meet the demand of these applications.

One method of acceleration is to divide the entire computation into offline and online processes. In the offline process, a set of models are computed and the results are stored in the memory or computer hard disk. Then in the online process, solutions can be interpolated from the pre-computed results. These methods include the model order reduction \cite{Wilhelmus2008}, the characteristic basis function \cite{prakash2003characteristic}, the reduced basis method \cite{Noor1980, dang2017quasi}, and etc. The idea of these schemes is to pay more memory in return for faster speed. Moreover, artificial neural network has also been used to optimize circuit\cite{zaabab1995neural} and accelerate the design of RF and microwave components \cite{Zhang2000}\cite{zhang2003artificial}. However, the extension capability is still limited for most of these methods, and they are mainly used to describe systems with few parameters.

With rapid development of big data technology and high performance computing, deep learning methods have been applied in many areas and significantly improve the performance of voice and image processing \cite{Hinton2006, LuCun2015}. These dramatic improvements rely on the strong approximation capability of deep neural networks. Recently, researchers have applied the deep neural networks to approximate complex physical systems \cite{Ehrhardt2017}\cite{lerer2016learning}, such as fluid dynamics \cite{Tompson2016, Guo2016}, Schr\"{o}dinger equations \cite{Mills2017} and rigid body motion\cite{Byravan2017SE3}. In these works, the deep neural networks "learn" from data simulated with traditional solvers. Then it can predict field distribution in a domain with thousands or millions unknowns. Furthermore, it has also been applied in capacitance extraction with some promising results \cite{yao2016machine}. The flexility in modeling different scenarios is also significantly improved compared with traditional techniques using artificial neural networks.

In this study, we investigate the feasibility of using deep learning techniques to accelerate electromagnetic simulation. As a starting point, we aim to compute 2D or 3D electric potential distribution by solving the 2D or 3D Poisson's equation. We extended the deep neural network structure in \cite{Tompson2016} and proposed a
approximation model based on fully convolutional network \cite{long2015fully}. We apply L2 regularization\cite{ng2004feature} in the objective function in order to prevent over-fitting and improve the prediction accuracy.
In the offline training stage, a finite-difference solver is used to model inhomogeneous permittivity distribution and point-source excitation at different locations, the permittivity distribution, excitation, and potential field are used as training data set. The input data include the permittivity distribution and the location of excitation, the output data is the electric potential of the computation domain. Then in the online stage, the network can mimic the solving process and correctly predict the electric potential distribution in the domain. Different from traditional algorithms, the method proposed in this paper is an end-to-end simulation driven by data. The computational complexity of the network is fixed and much smaller than that of traditional algorithms, such as the finite-difference method. Preliminary numerical studies also support our observations.

This paper is organized as follows: In Section 2 we introduce the data model and the deep convolutional neural network model used in the computation. In Section 3 we show more details of preliminary numerical examples and compare the accuracy and computing time with the algorithm using finite-difference method. Conclusions are drawn in Section 4.

\section{Formulation}
\subsection{Finite Difference Method Model}
The electrostatic potential in the region of computation with Dirichlet boundary condition can be described as
\par
\begin{equation}
\nabla\cdot(\varepsilon(\v{r})\nabla\phi(\v{r}))=-\rho(\v{r}) \,,
\label{eq10}
\end{equation}
\begin{equation}
\phi|_{\partial D}=0 \,,
\label{eq20}
\end{equation}
\noindent where $\phi(\v{r})$ is the electric potential in Domain $D$, $\rho(\v{r})$ represents distribution of electric charges, and $\varepsilon(\v{r})$ represents dielectric constant. \eq{eq20} describes the Dirichlet boundary condition, which enforces the value of potential to be zero along the boundary.
\par
The above equations are solved using the finite difference method. The domain of computation is partitioned into subdomains using Cartesian grids. The electric potential and electric charge density in each subdomain is assumed constant. Central difference scheme is used to approximate the derivative in \eq{eq10}. If the computation domain is 2D, then we can write \eq{eq10} as
\begin{equation}
\begin{split}
&\frac{\varepsilon_{i+\frac{1}{2},j}\frac{\phi_{i+1,j}-\phi_{i,j}}{\Delta x}-\varepsilon_{i-\frac{1}{2},j}\frac{\phi_{i,j}-\phi_{i-1,j}}{\Delta x}}{\Delta x}+\\
&\frac{\varepsilon_{i,j+\frac{1}{2}}\frac{\phi_{i,j+1}-\phi_{i,j}}{\Delta y}-\varepsilon_{i,j-\frac{1}{2}}\frac{\phi_{i,j}-\phi_{i,j-1}}{\Delta y}}{\Delta y}=-\rho_{i,j}
\end{split}
\label{eq30}
\end{equation}
and
\begin{equation}
\varepsilon_{i+a\frac{1}{2},j+a\frac{1}{2}}=\frac{\varepsilon_{i,j}+\varepsilon_{i+a,j+b}}{2} , a \in \{-1, 1\}
\label{eq40}
\end{equation}
where $(i,j)$ represents the location of the subdomain in the grid.
if the comptation domain is 3D, then \eq{eq10} can be written as
\begin{equation}
\begin{split}
&\frac{\varepsilon_{i+\frac{1}{2},j,k}\frac{\phi_{i+1,j,k}-\phi_{i,j,k}}{\Delta x}-\varepsilon_{i-\frac{1}{2},j,k}\frac{\phi_{i,j,k}-\phi_{i-1,j,k}}{\Delta x}}{\Delta x}+\\
&\frac{\varepsilon_{i,j+\frac{1}{2},k}\frac{\phi_{i,j+1,k}-\phi_{i,j,k}}{\Delta y}-\varepsilon_{i,j-\frac{1}{2},k}\frac{\phi_{i,j,k}-\phi_{i,j-1,k}}{\Delta y}}{\Delta y}+\\
&\frac{\varepsilon_{i,j,k+\frac{1}{2}}\frac{\phi_{i,j,k+1}-\phi_{i,j,k}}{\Delta z}-\varepsilon_{i,j,k-\frac{1}{2}}\frac{\phi_{i,j,k}-\phi_{i,j,k-1}}{\Delta z}}{\Delta z}=-\rho_{i,j}
\end{split}
\label{eq30}
\end{equation}
and
\begin{equation}
\varepsilon_{i+a\frac{1}{2},j+b\frac{1}{2},k+c\frac{1}{2}}=\frac{\varepsilon_{i,j,k}+\varepsilon_{i+a,j+b,k+c}}{2} , a,b,c \in \{-1,0,1\}
\end{equation}
where $(i,j,k)$ represents the location of the subdomain in the grid.
The above equation in each subdomain construct a linear system of equations $\bar{\bar{A}}\cdot\vg{\bar{\phi}} = - \vg{\bar{\rho}}$, where $\bar{\bar{A}}$ is symmetric and positive semi-definite. LU decomposition or conjugate gradient method can be applied to solve this equation.
\par
\begin{figure}
	\centering
	\includegraphics[height=3cm]{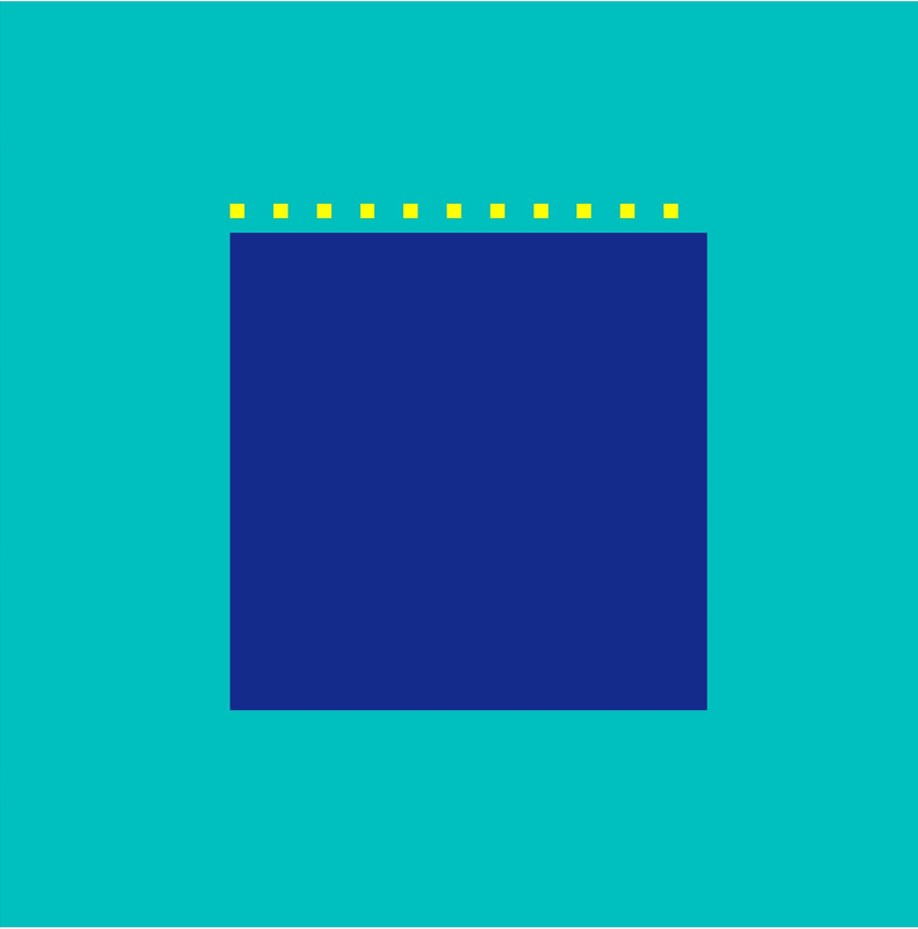}
	\caption{2D Modeling Setup: Yellow points are the 11 positions for source , blue area is where we predict}
	\label{2dmodel}
\end{figure}
\par
\begin{figure}
	\centering
	\includegraphics[height=4cm]{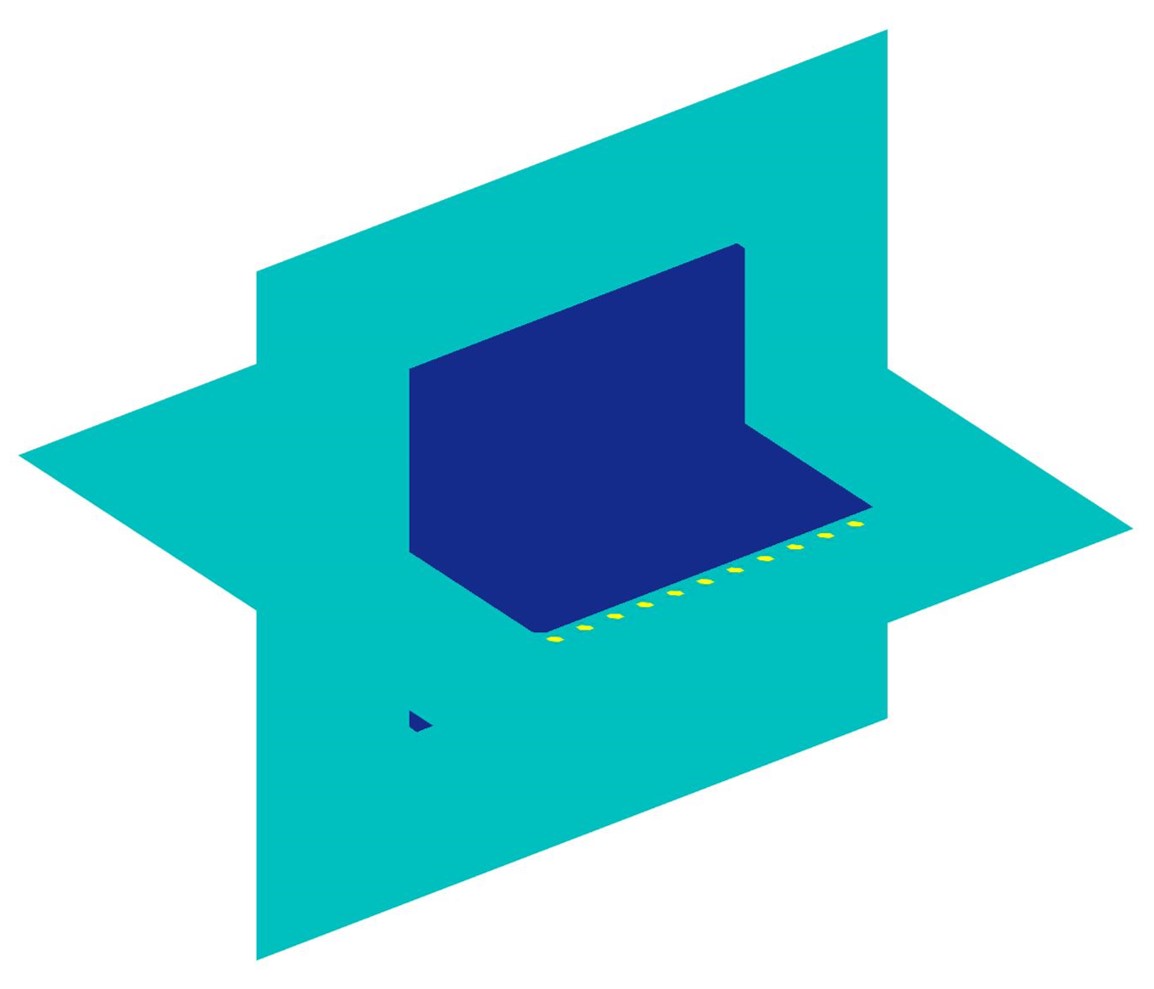}
	\caption{3D Modeling Setup: Yellow points are the 11 positions for source , blue area is where we predict}
	\label{3dmodel}
\end{figure}
\begin{figure}[!t]
	\centering
	\foreach \t in {1,...,5}{	\subfigure
		{		\includegraphics[width=25mm]{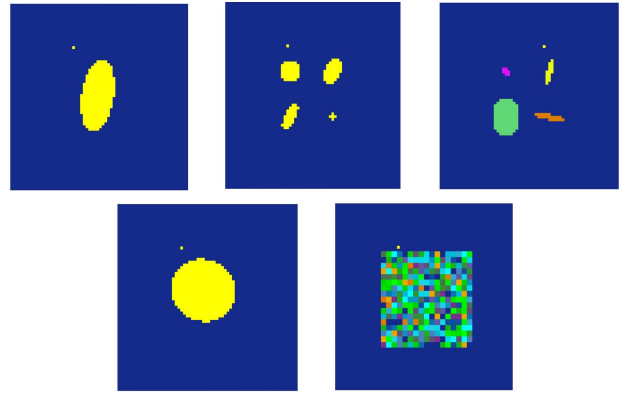}		}}
	\caption{Five examples of five scenarios of 2D model: excitation source and dielectric constant distribution}
	\label{2D example}
\end{figure}
\begin{figure}[!t]
	\centering
	\foreach \t in {1,...,3}{	\subfigure
		{		\includegraphics[height=2.6cm]{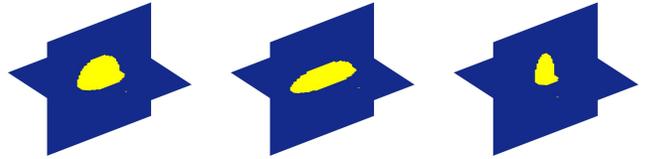}		}}
	\caption{Three examples of 3D model: excitation source and dielectric constant distribution}
	\label{3D example}
\end{figure}

\begin{figure*}
	\centering
	\includegraphics[width=12.7cm,height=8cm]{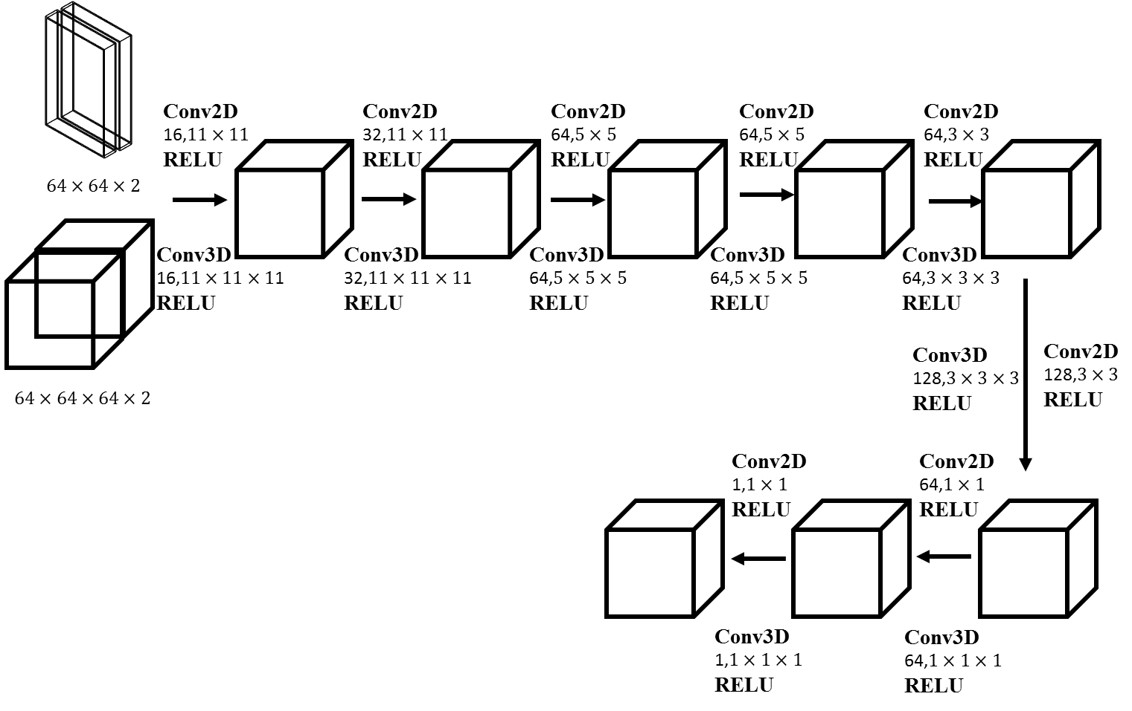}
	\caption{Convolutional nueral network for solving Possion's equation}
	\label{Network Model}
\end{figure*}

\subsection{ConvNet model}
Neural networks have excellent performance for function fitting. One can approximate a complex function using powerful function fitting method based on deep neural networks \cite{Tompson2016}.
Convolutional neural networks (CNN) have excellent performance in learning geometry and predicting per-piexl in images\cite{lawrence1997face}\cite{krizhevsky2012imagenet} and fully convolutional networks have been proven powerful models of pixelwise prediction.
In this paper, the problem that we model has various locations of the excitation and dielectric constant distribution, so the geometric characteristics of this problem are obvious. They all need to be considered in the design of the network layers.
Therefore, the input of the network includes distribution of electrical permittivity and the source information.
The electrical permittivity distribution is represented as a two-dimensional or three-dimensional array with every element $(i,j)$ or $(i,j,k)$ represents electrical permittivity at grid $(i,j)$ or $(i,j,k)$. The source information is also represented by a two-dimensional or three-dimensional array, in which every element represents the distance between the source and a single grid. The distance function can provide a good representation of the universial source information and if the case is 2D, then the distance function can be written as
\par
\begin{equation}
f(i,j)_n=\sqrt{(i-i_n )^2+(j-j_n )^2}, n \in \{1,2...,11\}
\label{eq50}
\end{equation}
where $i$,$j$ is the location of grids in the predicted area and $i_n$,$j_n$ is source excitation's location that has 11 different positions.
if the case is 3D, then the distance function can be written as
\par
\begin{equation}
\begin{split}
	&f(i,j,k)_n=\sqrt{(i-i_n )^2+(j-j_n )^2+(k-k_n )^2},\\
	& \quad\quad\quad\quad \quad\quad\quad\quad  \quad\quad\quad\quad \quad\quad\quad\quad  n \in \{1,2...,11\}
	\end{split}
\end{equation}
where $i$,$j$,$k$ is the location of grids in the predicted area and $i_n$,$j_n$,$k_n$ is source excitation's location that has 11 different positions.

The setup of deep neural network is based on optimization process that adjusts the network parameters to minimize the difference between "true" values of function and the one predicted by the network on a set of sampling points. In this problem, the loss function in optimization is defined to measure the difference between the logarithm of predicted potential and the one obtained by FDM, it can be written as
\par
\begin{equation}
loss_{obj}=\|log_{10}(\phi)-log_{10}(\widehat{\phi}) \|^2\,,
\end{equation}
and L2 regularization expression is included in the cost function to prevent over-fitting, the final cost function is written as:
\begin{equation}
f_{obj}=loss_{obj}+ \frac{\lambda}{2n}\sum_{w}w^{2}\,,
\label{eq60}
\end{equation}
where $\phi$ is the predicted potential, $\widehat{\phi}$ is the potential solved by FDM, $\lambda$ is a hyperparameter, \emph{n} is the amount of training samples, \emph{w} is weights of the network. The use of logarithm of potential is to avoid the instability in optimization due the fast attenuation in the distribution of electrical potential. It can also help to improve the accuracy of prediction.
L2 regularization implies a tendency to train as small a weight as possible and retain a larger weight that can be guaranteed to significantly reduce the original loss.
The internal structure of fully convolution neural network model is shown in the \fig{Network Model}. It consists of seven stages of convolution and Rectifying Linear layers (ReLU)\cite{glorot2011deep} but there are no pooling layers beacuse features in this problem are not complicated. The input data of ConvNet model includes the permittivity distribution and location of excitation expressed as the distance function, the output data is the predicted electric potential of the computation domain.
\par
\begin{figure}
	\centering
	\subfigure{
		\begin{minipage}{3cm}
			\centering
			\includegraphics[width=2.8cm]{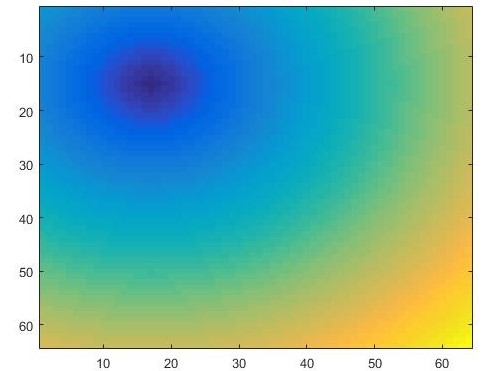}
		\end{minipage}
	}
	\subfigure{
		\begin{minipage}{3cm}
			\centering
			\includegraphics[width=2.64cm]{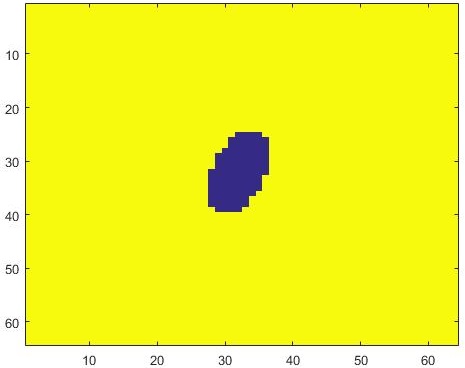}
		\end{minipage}
	}
	\caption{64$\times$64 Input: distance matrix and premittivity distribution matrix}
	\label{2D excitaotion and dielectric constant matrix}
\end{figure}
\par
\begin{figure}
	\centering
	\subfigure{
		\begin{minipage}{4cm}
			\centering
			\includegraphics[width=4cm]{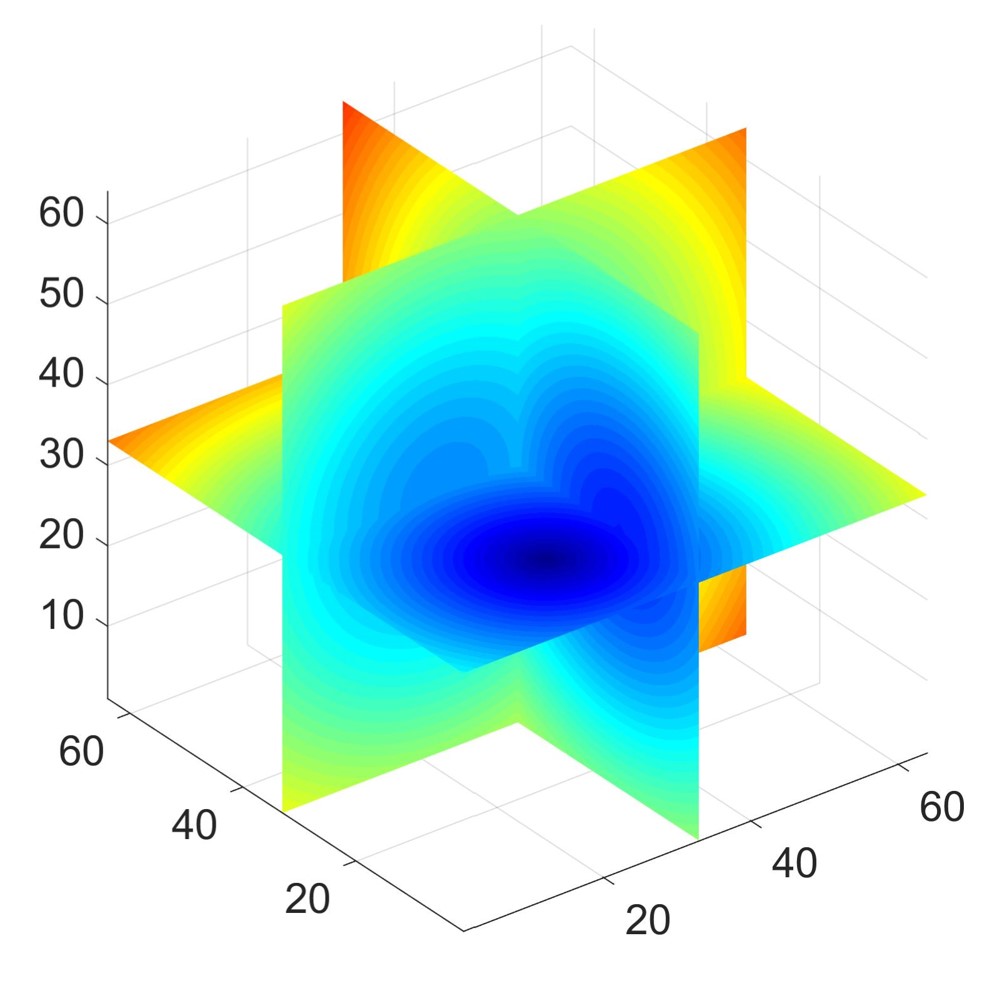}
		\end{minipage}
	}
	\subfigure{
		\begin{minipage}{4cm}
			\centering
			\includegraphics[width=4cm]{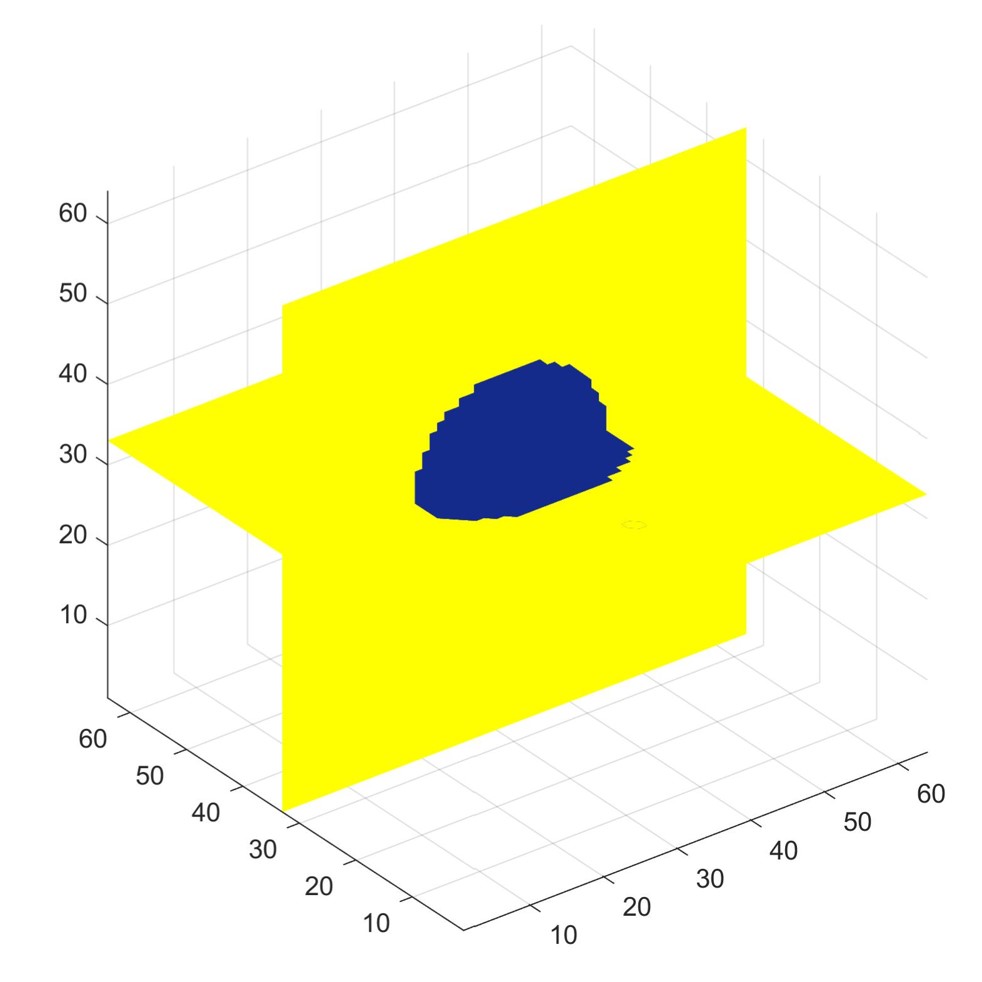}
		\end{minipage}
	}
	\caption{64$\times$64$\times$64 Input: distance matrix and premittivity distribution matrix}
	\label{3D excitaotion and dielectric constant matrix}
\end{figure}
\begin{figure*}[!t]
	\centering
	\foreach \t in {1,...,3}{	\subfigure
		{		\includegraphics[width=50mm]{2dreslut1\t.jpg}		}}
	\par
	One result of scenario 1: \emph{Left}: FDM result, \emph{Midlle}: ConvNet result, \emph{Right}: error distribution
	\par
	\centering
	\foreach \t in {1,...,3}{	\subfigure
		{		\includegraphics[width=50mm]{2dreslut2\t.jpg}		}}
	\par
	One result of scenario 2: \emph{Left}: FDM result, \emph{Midlle}: ConvNet result, \emph{Right}: error distribution
	\par
	\centering
	\foreach \t in {1,...,3}{	\subfigure
		{		\includegraphics[width=50mm]{2dreslut3\t.jpg}		}}
	\par
	One result of scenario 3: \emph{Left}: FDM result, \emph{Midlle}: ConvNet result, \emph{Right}: error distribution
	\par
	\centering
	\foreach \t in {1,...,3}{	\subfigure
		{		\includegraphics[width=50mm]{2dreslut4\t.jpg}		}}
	\par
	One result of scenario 4: \emph{Left}: FDM result, \emph{Midlle}: ConvNet result, \emph{Right}: error distribution
	\par
	\centering
	\foreach \t in {1,...,3}{	\subfigure
		{		\includegraphics[width=50mm]{2dreslut5\t.jpg}		}}
	\par
	One result of scenario 5: \emph{Left}: FDM result, \emph{Midlle}: ConvNet result, \emph{Right}: error distribution
	\par
	\caption{Results of 2D cases}
	\label{2Dresult}
\end{figure*}
\par
\section{Results And Analysis}
In this study, we solve the electrostatic problem in a square region (2D case) or a cube region (3D case).
In 2D cases, a square region is partitioned into $64\times 64$ grids, as shown in \fig{2dmodel}. The yellow points indicate the location of sampled excitation.
In 3D cases, a cube region is partioned into $64\times 64\times64$ grids, as shown in \fig{3dmodel}.The yellow points indicate the location of sampled excitation and the value of excitation is fixed at -10.
We aim to solve the potential field in the region of $32\times 32$ or $32\times32\times32$ colored in blue.
\par
In 2D case, we try 6 possible scenarios as shown in c\fig{2D example} and the background's permittivity of all scenarios is 1:
\subsection{scenario 1}
Scenario 1 has a single ellipse located in the center of the square region. This ellipse has different shapes whose semi-axis varies from 1 to 20 and rotation angle is randomly chosen between $\frac{\pi}{20}$ and $\pi$. The permittivity values of the target is randomly selected from [0.125,0.25,0.5,2,4,6].
\subsection{scenario 2}
Scenario 2 divides the square region into four identical parts and each part has a ellipse whose semi-axis varies from 1 to 8 and rotation angle is randomly chosen between $\frac{\pi}{20}$ and $\pi$. The four ellipses have different shapes but their permittivity valuse are the same and randomly selected from [0.125,0.25,0.5,2,4,6].
\subsection{scenario 3}
Scenario 3 divides the square region into four identical parts and each part has a ellipse whose semi-axis varies from 1 to 8 and rotation angle is randomly chosen between $\frac{\pi}{20}$ and $\pi$. The four ellipses have different shapes and their permittivity valuse are different and randomly selected from [0.125,0.25,0.5,2,4,6].
\subsection{scenario 4}
Scenario 4 has a single ellipse whose location moves in a small range. This ellipse has different shapes whose semi-axis varies from 1 to 12 and rotation angle is randomly chosen between  $\frac{\pi}{20}$ and $\pi$. The permittivity values of the target is randomly selected from [0.125,0.25,0.5,2,4,6].
\subsection{scenario 5}
scenario 5 has no special shapes and the predicted region is the region of $32\times 32$ colored in blue. The permittivity values of every four grids in the target is randomly chosen from 0.125 to 6.
\subsection{scenario 6}
Scenario 6 includes scenario 1 to 5.
\begin{figure*}[!t]
	\centering
	\foreach \t in {1,...,6}{	\subfigure
		{		\includegraphics[width=55mm]{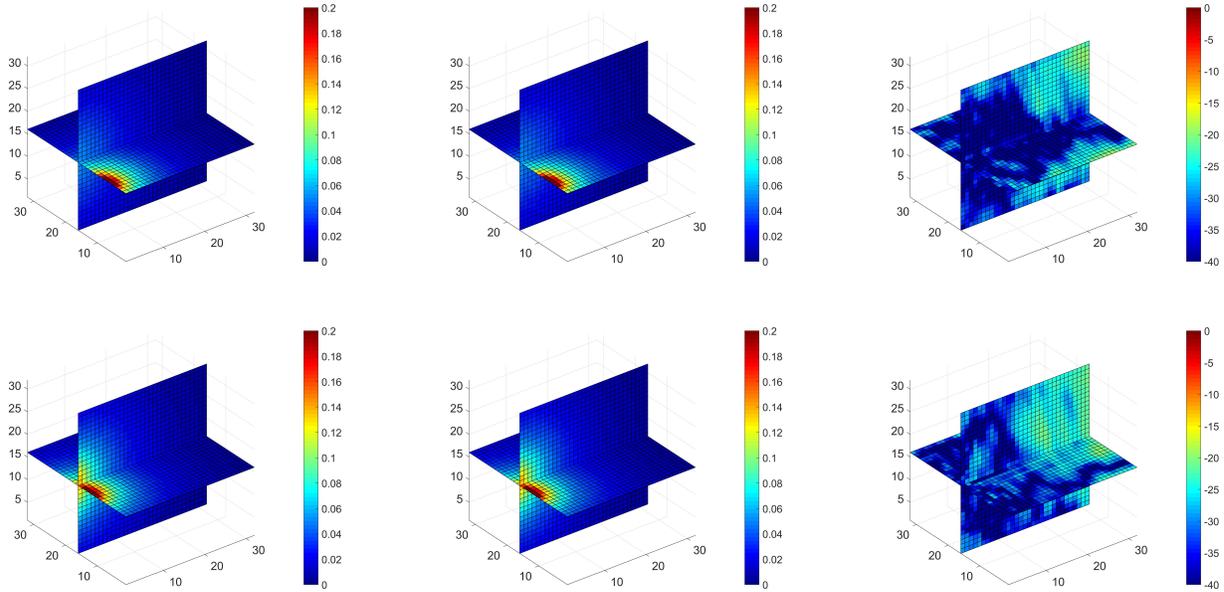}		}}
	\caption{Result of 3D cases: \emph{Left}: FDM result, \emph{Midlle}: ConvNet result, \emph{Right}: error distribution}
	\label{3Dresult}
\end{figure*}
\par
In the 3D case, ellipsoids with different shapes are located inside the predicted region. As shown in \fig{3D example}, their three semi-axis varies from 1 to 20.
\par
The convolution neural network takes two $64\times 64$ or $64\times64\times64$ arrays as input, as depicted in \fig{2D excitaotion and dielectric constant matrix}, \fig{3D excitaotion and dielectric constant matrix} and output is a $32\times32$ or $32\times32\times32$ array representing the field in the region of investigation. The training and testing data for the network are obtained by the finite-difference solver. For 2D cases,  We use 8000 samples for training and 2000 samples for testing in sceranio 1 to 5 and 40000 samples for training and 10000 samples for testing in scenario 6; for 3D cases, we use 4000 samples for training and 1000 samples for testing. The ConvNet model was implemented in Tensorflow and an Nvidia K80 GPU card is used for computation. The Adam\cite{kingma2014adam} Optimizer is used to optimize objective function in \eq{eq60}.


For more detailed comparison, we use relative error in the ConvNet model to measure the accuracy of the prediction. We first compute the difference between the ConvNet model predicted potential and the FDM generated potential. For a subdomain, the relative error is defined as
\begin{equation}
err(i,j)\ or\ err(i,j,k)=\frac{|\phi_{ConvNet}-\phi_{FDM}|}{\phi_{FDM}} \,,
\end{equation}
where $\phi_{ConvNet}$ and $\phi_{FDM}$ are predicted and "true" potential field, respectively.
The average relative error of the $n$-th testing case is the mean value of relative error in all subdomains:
\begin{equation}
err_{aver_n}=20\lg10(\frac{\sum_{i}\sum_{j}err(i,j)}{\sum_{i}\sum_{j} 1 }), for\ 2D\ case
\end{equation}
\begin{equation}
err_{aver_n}=20\lg10(\frac{\sum_{i}\sum_{j}\sum_{k}err(i,j)}{\sum_{i}\sum_{j}\sum_{k} 1 }), for\ 3D\ case
\end{equation}
\par
\fig{2Dresult} shows one result of scenario 1 to 5 in 2D cases, which is randomly chosen from the testing samples. It can be observed that the predicted potential field distribution agrees well with the one computed by finite difference method. The final average relative error of the prediction in scenario 1 to 5 by ConvNet model is -41dB, -41dB, -38dB, -38dB, -40dB respectivly. And in the scenario 6, the final average relative error of the prediction  is -38dB. The proposed ConvNet model shows a good prediciton capability and good generalization ability for 2D cases.
\begin{figure}
	\centering
	\includegraphics[width=6cm]{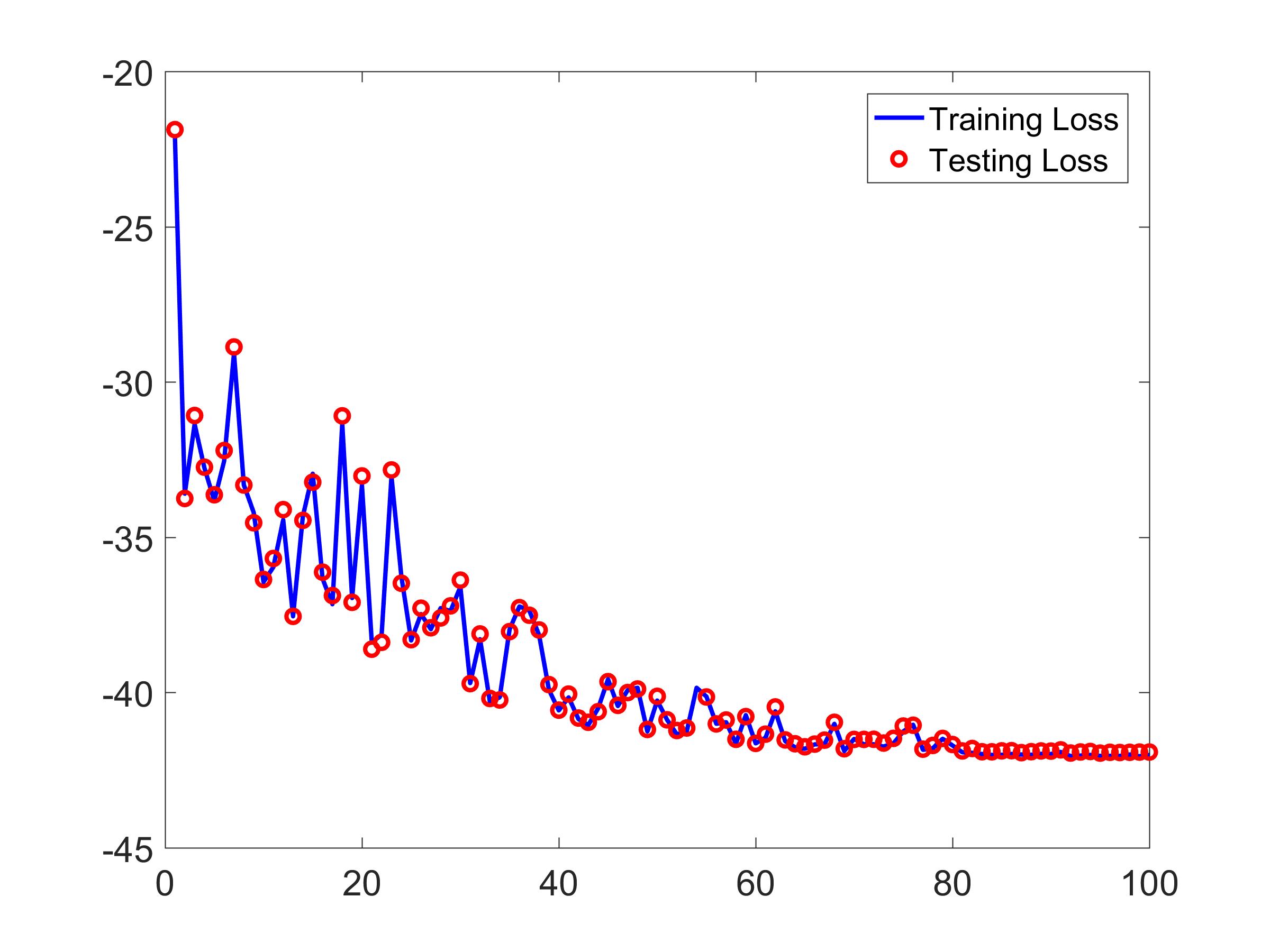}
	\caption{Loss curves of trainning and testing in 2D cases }
	\label{traintestloss}
\end{figure}
The result of 3D cases is visualized in \fig{3Dresult}. The difference between ConvNet model's prediction and FDM results is little and the final average relative error is -31dB. The ConvNet model can do good predictions on 3D cases which are more complicated than 2D cases. The good prediction capability and generalization ability of proposed ConvNet model is verified.
\par
\fig{traintestloss} shows that in 2D cases, the curve of testing loss agrees well with training loss's curve, which means the ConvNet model do not over-fit the training data.

Using this model, the CPU time is reduced significantly for 2D cases and 3D cases. For example, using FDM to obtain 2000 sets of 2D potential distribution takes 16s but using ConvNet model only takes 0.13s, and using FDM to obtain 5 sets of 3D potential distribution takes 292s but using ConvNet model only takes 1.2s. This indicates the possibility to build a realtime electromagnetic simulator.

\section{Conclusion}
In this study, we investigate the possibility of using deep learning techniques to reduce the computational complexity in electromagnetic simulation. Here we compute the 2D amd 3D electrostatic problem as an example. By building up a proper convolutional neural network, we manage to correctly predict the potential field with the average relative error below 1.5\% in 2D cases and below 3\% in 3D cases. Moreover, the computational time is significantly reduced. This study shows that it may be possible to take advantage of the flexibility in deep neural networks and build up a fast electromagnetic solver that may provide realtime responses. In the future work, we will further improve the accurracy of 3D cases' preidiction and try to build a fast electromagnetic realtime simulator.





\end{document}